\documentclass[a4paper,12pt]{article}

\usepackage[utf8]{inputenc}
\usepackage[dvips]{color,graphicx}
\usepackage{mathrsfs}
\usepackage{amssymb}
\usepackage{amsmath}
\usepackage{feynmp}
\usepackage{slashed}
\usepackage{latexsym}
\usepackage{graphicx}
\usepackage{verbatim}
\usepackage[normalem]{ulem}

\newcommand{\be}{\begin{equation}}
\newcommand{\ee}{\end{equation}}
\newcommand{\ba}{\begin{array}}
\newcommand{\ea}{\end{array}}
\newcommand{\bea}{\begin{eqnarray}}
\newcommand{\eea}{\end{eqnarray}}
\newcommand{\balg}{\begin{align}}
\newcommand{\ealg}{\end{align}}
\newcommand{\bit}{\begin{itemize}}
\newcommand{\eit}{\end{itemize}}
\newcommand{\trm}[1]{\textrm{#1}}
\newcommand{\mbf}[1]{\mathbf{#1}}

\newcommand{\mbb}[1]{\mathbb{#1}}
\newcommand{\msc}[1]{\mathscr{#1}}

\newcommand{\Mpc}{\trm{\Mpc}}
\newcommand{\yr}{\trm{\yr}}
\newcommand{\eV}{\trm{\eV}}

\newcommand{\nn}{\nonumber}

\newcommand{\vtw}{\vspace{.2cm}}

\newcommand{\vsx}{\vspace{.6cm}}

\newcommand{\tr}[1]{\textrm{Tr}[#1]}

\begin{document}

\title{
\Large \bf
Lepton mixing and discrete symmetries}
\author{
{D. Hernandez$^{a}$\thanks{email:
\tt dhernand@ictp.it}~~\,and
\vspace*{0.15cm} ~A. Yu. Smirnov
$^{a}$\thanks{email:
\tt smirnov@ictp.it}
} \\
{\normalsize\em $^{a}$The Abdus Salam International Centre for Theoretical    
Physics} \\
{\normalsize\em Strada Costiera 11, I-34014 Trieste, Italy 
\vspace*{0.15cm}}
}
\date{}
\maketitle
\thispagestyle{empty}
\vspace{-0.8cm}
\begin{abstract}
The pattern of lepton mixing can emerge  from breaking a flavor symmetry in different ways in the 
neutrino and charged lepton Yukawa sectors.  In this framework, we derive the model-independent 
conditions imposed on  the mixing matrix by the structure of discrete   groups of the von Dyck type 
which include ${\bf A}_4$,  ${\bf S}_4$ and ${\bf A}_5$. We show that, in general, these  conditions 
lead to at least two equations for the  mixing  parameters (angles and CP phase $\delta$). These  
constraints, that correspond to unbroken residual  symmetries, are consistent with non-zero 13 
mixing  and deviations from maximal 2-3  mixing. For the simplest case, 
 which leads to an $\mbf{S}_4$ model  and  reproduces the allowed values of 
the mixing angles, we predict $\delta = (90^\circ - 120^{\circ})$.
\end{abstract}

\vspace{1.cm}
\vspace{.3cm}

\newpage


\section{Introduction}


The fact that the leptonic mixing angles seem  to take ``special'' values \cite{fit} has long been 
presented as an argument for the existence of  an approximate underlying symmetry. Based on 
this, the following program for model building has been devised \cite{altarelli-feruglio}. 
One starts with a Lagrangian symmetric under a flavor group $G$ and breaks this symmetry down to two different symmetries described by subgroups, $G_\nu$ and  $G_\ell$, that are preserved in the neutrino 
and the charged lepton sector respectively. 
These different residual symmetries of  the Yukawa sectors are responsible for the mismatch of the 
rotations which diagonalize the neutrino  and charged lepton matrices. In particular, they 
can give rise to the tribimaximal (TBM) \cite{tbm} and bimaximal (BM) \cite{bm} mixing patterns.
Inversely, the  flavor group $G$ can be reconstructed from the residual symmetries of the mass 
matrices. This constructive approach was pioneered in \cite{lam} 
where it was shown that the minimal $G$ that included all the symmetries of 
the neutrino  mass matrix and leads to TBM mixing is ${\bf S}_4$.  
In \cite{Ge:2011qn,Ge:2011ih,Ge:2010js,He:2012yt} 
consequences of the postulated residual symmetries of the neutrino mass matrix 
for the lepton mixing have been explored.

In this paper we reformulate and generalize  the described approach. We deviate from the road 
pursued in \cite{lam} by recognizing that  not all the symmetries of the neutrino mass term 
necessarily belong to $G$ since they appear accidentally anyway~\cite{grimus}.
We will assume in this paper that the  subgroup of $G$ conserved by the neutrino mass term is 
only a $\mbf{Z}_2$. The consequence of this milder assumption is that the mixing matrix is not 
completely fixed thus broadening greatly the possibilities.

It is known that in specific models different symmetry groups  $G$ can lead to the same mixing. 
Inversely, the same symmetry group $G$ can  lead to different mixing patterns, e.g. TBM or BM. Such 
an  ambiguity makes rather unclear the  connection between symmetry groups and specific mixing 
patterns. 
In this paper we  develop a model-independent  formalism which  clarifies a correspondence between 
symmetries and mixing pattern and establishes   to which extent non-abelian discrete symmetries can 
fix the mixing. 

The paper is organized as follows.  In Sec.~\ref{sym-building}  we reformulate the program of  
``symmetry building''. 
We conclude that the relevant symmetries are in general of the von Dyck type.   
We obtain the condition which the mixing matrix 
should satisfy in terms of symmetry generating elements. In Sec.~\ref{arbitrary}.  
using this condition we derive the relations between the mixing angles that appear 
from different symmetries.  Sec.~\ref{discussion}  contains discussion and conclusions.


\section{Mixing and symmetry building} \label{sym-building}


We assume throughout this paper that the neutrino masses  are of Majorana type (the Dirac case 
allows for a broader set of symmetries and will be  considered separately). We begin by writing the 
corresponding Lagrangian in the mass basis  
\be
\msc{L} = \frac{g}{\sqrt{2}}\bar{\ell}_LU_{PMNS}\gamma^\mu \nu_L W^+_{\mu}  + \bar{E}_R m_\ell \ell_L + 
\frac{1}{2}  \bar{\nu^c}_L m_\nu \nu_L + \dots  \label{lag} + \trm{h.c.} \,,
\ee
where $U$ is the mixing matrix,  $\ell = (e,\, \mu,\, \tau)^T$, $\nu = (\nu_1,\,\nu_2,\,\nu_3)^T$ are vectors of mass states and $m_\ell$, $m_\nu$ are the diagonal mass matrices   of the charged leptons and neutrinos. 
  
Taken by itself, each mass term in Eq.~\eqref{lag} has certain symmetries. 
The neutrino mass term has two independent  and non-trivial $\mbf{Z}_2$ symmetries with respect to which 
$\nu$ transforms as
\be
\nu  \rightarrow S_i\nu \,,\quad m_\nu \rightarrow S^T_im_\nu S_i = m_\nu \,,\quad k = 1,\,2
\ee
with
\be
S_1 = \left( \begin{array}{ccc}
1 && \\
& -1 & \\
&& - 1
\end{array} \right)  \,, \quad S_2 = \left( \begin{array}{ccc}
-1 && \\
& 1 & \\
&& -1
\end{array} \right)  \,. \label{S-matrices}
\ee
The mass matrix is also invariant under an additional transformation $S_3$ described by
\be
S_3 = S_1S_2 =  \left( \begin{array}{ccc}
-1 && \\
& -1 & \\
&& 1
\end{array} \right)\,.
\ee
For simplicity, the matrices $S_i$ are taken to be  elements of $SU(3)$, so they satisfy 
$\trm{Det}[S_i] = 1$.

\vtw
The charged lepton mass term has a  full $U(1)^3$ symmetry.  Restricting to discrete symmetries 
implies that the values of the phases must be rational  divisors of  $2\pi i$ and the symmetry that 
remains is $\mbf{Z}_m$. The transformation can be written explicitly as
\be
\ell \rightarrow T \ell_L \,,\quad \ell_R \rightarrow T \ell_R\,, \label{cl-transf}
\ee
where
\be
T =  \left( \begin{array}{ccc}
e^{2\pi i\frac{k_1}{m}} && \\
& e^{2\pi i\frac{k_2}{m}} & \\
&& e^{2\pi i\frac{k_3}{m}}
\end{array} \right)  \label{Tdef} 
\ee 
satisfies $T^m = \mbb{I}$. 

Once again, we assume that $T$ belongs to $SU(3)$ so that  $\trm{Det}[T]=1$ which implies
\be
k_1+k_2+k_3 =  0,\, m,\,2m \,. \label{ass1}
\ee

We take one of the $k_i$ to be zero, which fixes  the $\mbf{Z}_m$ charges of the charged leptons, 
and consider in the rest of the paper the three $T$ matrices
\begin{align}
T_e & = \trm{diag}\{ 1, e^{2\pi i k/m}, e^{-2\pi i k/m} \} \,, \\
T_\mu & =  \trm{diag}\{ e^{2\pi i k/m}, 1, e^{-2\pi i k/m} \} \,, \\
T_\tau & =  \trm{diag}\{ e^{2\pi i k/m}, e^{-2\pi i k/m}, 1 \} \,. 
\end{align}
The assumption that both $S_i$ and $T$ are elements  of $SU(3)$ simplifies our considerations, 
e.g., it reduces the number of independent phases of $T$.
A more general form for the $T$ generator will be discussed elsewhere \cite{prep}.  

\vtw
Notice that the charged current interactions in Eq.~\eqref{lag}  are not invariant under the 
transformations $S_i$ or $T$ in this basis.

\vtw
We now examine these symmetries in an arbitrary basis, related to the mass basis by the rotations
\be
U_\nu\nu_L = \nu_L' \,,\quad U_\ell\ell_L = \ell_L' \,, 
\quad V_\ell \ell_R = \ell'_R
\ee
The mass matrices of neutrinos and charged leptons become 
\be
 m_{\nu U} = U^*_\nu m_\nu U^\dagger_\nu \,,\quad  m_{\ell U} =  V_\ell m_\ell U^\dagger_\ell \,.
\ee
It is easy to see that $m_{\nu U}$ is invariant under the transformed matrix $S_{iU}$
\be
S_{iU} = U_\nu S_iU_\nu^\dagger , 
\label{ST-transforms}
\ee
while $m_{\ell U}$ is invariant under the transformation
\be
m_{\ell U} = T_{\alpha V}^\dagger m_{\ell U} T_{\alpha U} \,
\ee
where
\be
T_{\alpha V} = V_\ell^\dagger T_{\alpha} V_\ell \,,\quad T_{\alpha U} = U_\ell^\dagger T_{\alpha}U_\ell \,.
\ee
These satisfy the same $\mbf{Z}_2$ and $\mbf{Z}_m$ relations as in the mass basis\footnote{The mismatch between $T_{\alpha U}$ and $T_{\alpha V}$ 
is merely the mismatch between the bases of $\ell_L$ and $\ell_R$ since there exists a basis, the mass basis, in which they are the same. They correspond nevertheless to the same group element.} 
\be
S^2_{iU} = T^m_{\alpha V} = T^m_{\alpha U} = \mbb{I} \,. 
\label{rel1}
\ee
The transformations of the RH components of leptons are  irrelevant for the discussion of the mixing 
matrix $U_{PMNS}$, since only the left-handed lepton  basis enters in the physical mixing. The 
neutrino and charged lepton mass  terms in an arbitrary  basis are invariant under the $\mbf{Z}_2$ 
and $\mbf{Z}_m$ symmetries irrespective of what the mixing matrix is. 
In the new basis the weak interaction term reads
\be
\bar{\ell}_LU_\ell U_{PMNS}U_\nu^\dagger\nu_L. 
\ee
It becomes diagonal, and therefore invariant under the flavor symmetry 
transformations in Eq.~\eqref{ST-transforms}, if  
\be
U_\ell^\dagger U_\nu = U_{PMNS}. 
\label{pmns}
\ee

The first relation in Eq.~\eqref{ST-transforms}  can be rewritten as 
\be
S_{iU} U_\nu  = U_\nu S_i.  
\label{ST-transforms1}
\ee
It means that the columns of the matrix $U_\nu$, which coincides 
with $U_{PMNS}$ in the flavor basis ($U_\ell = \mbb{I}$), are the 
eigenstates of the symmetry matrix $S_{iU}$ with eigenvalues 
given by the diagonal elements of $S_i$. If $S_{iU}$ is known (postulated), 
 Eq.~\eqref{ST-transforms1} can be used to put conditions on 
the elements of mixing matrix.

\vtw
We now make the key assumption that the residual symmetries in both the charged lepton and neutrino 
sectors are one-generator  groups. That is,  we assume that $\{S_{iU},\,T_{\alpha U}\}$  form  a set 
of generators for the flavor  group $G$ for given $i$ and $\alpha$ (this excludes in particular, the case in which the residual symmetry in the charged lepton sector is a Klein group). Then the definition of $G$ requires a relation linking  $S_{iU}$ and $T_{\alpha U}$. In this paper we will study
\be
(S_{iU}T_{\alpha U})^p = ( U_{PMNS}S_iU_{PMNS}^\dagger T_\alpha)^p = \mbb{I} \,
\label{rel2}
\ee
which, for small values of $m$ and $p$,
leads to well known groups.  Other relations with the structure $W^p= \mbb{I}$, where $W$ is a product of $S_{iU}$ and $T_{\alpha U}$  matrices are possible. 
We stress that all these relations, including Eq.~\eqref{rel2}, do not depend on $U_\nu$ or $U_\ell$ as such but only in the combination $U_{PMNS}$ in  Eq.~\eqref{pmns}. Thus, the mixing pattern, which follows from symmetry, is independent of the 
basis,   as it should be. 

Notice that Eq.~\eqref{cl-transf}  doesn't imply that $\ell_L$ and $\ell_R$ must transform under the 
same representation of $G$. The requirement  that a cyclic group is preserved in the charged lepton 
sector is enough to guarantee that  Eq.~\eqref{cl-transf} holds. 

Since we use only one of $S_{iU}$  as a generator 
of the flavor group $G$, the others $S_{iU}$ that do not belong to $G$ should be treated as 
generators of \emph{accidental symmetries} of the mass matrix (recall that all three symmetries  $S_{iU}$ exist simply 
because the Majorana mass matrix is  symmetric). This point is 
the departure from \cite{lam} where it was assumed that all $S_i$  belong to $G$. In our case, as we will see later, the choice of $S_i$ leads  to different relations between the mixing parameters.
The accidental symmetries play no role in our model-independent considerations. 
In specific models, the situation  with a single $\mbf{Z}_2$ in the neutrino sector can be 
realized in two different ways: (a) from  the beginning the flavor group contains only one generator  
$\mbf{Z}_2$ or (b) the flavor group contains  both generators, but one is broken in the lowest 
order. The inclusion of the second $\mbf{Z}_2$  into $G$  enhances the symmetry and therefore leads 
to additional constraints on the mixing parameters.  In this sense, the framework explored in this 
paper is more general than the one in \cite{lam}.

The relation in Eq.~\eqref{rel2}  can be presented in a different form. We consider the combination $W_U = T_{\alpha U}^{m-1}S_{iU}$, so that 
\be
S_{iU}T_{\alpha U}W_U = \mbb{I} \label{rel5}\,.
\ee
Then, noticing that $T^{m -1}_{\alpha U}S_{iU}=(S_{iU} T_{\alpha U})^{-1}$, one finds from  Eq.~\eqref{rel2}
\be
W^p_U = (T_{\alpha U}^{m-1}S_{iU})^p= [(S_{iU} T_{\alpha U})^{-1}]^p = \mbb{I} \label{rel3} \,.
\ee
The three relations 
\be
S_{iU}^2 = T_{\alpha U}^m = W_U^p = S_{iU}T_{\alpha U}W_U = \mbb{I}
\label{rel4} 
\ee
define the von Dyck group $D(2,m,p)$ \cite{johnson}. Some well known cases are
\begin{align}
D(2,2,3) & = {\bf S}_3 \,, \nn\\ 
D(2,3,3) & = {\bf A}_4 \,, \nn\\
D(2,3,4) & = {\bf S}_4 \,, \label{finite}\\
D(2,3,5) & = {\bf A}_5  \,. \nn
\end{align}
We will take from now on the relations  in Eq.~\eqref{rel2} and 
(\ref{rel1})  as the defining relation for the von Dyck group $D(2,m,p)$. 

One ambiguity appears at Eq.~\eqref{finite}.  An $m \leftrightarrow p\,$ exchange defines the same von 
Dyck group despite requiring different $T_{\alpha U}$  matrices. Starting from $G$, different 
$T_{\alpha U}$ matrices may appear as a result of the  choice of the representations for the fields 
and of the ways of breaking the symmetry. In such  cases \emph{the same group may lead to different 
patterns of mixing}. 
We stress again that the fact that we consider Majorana masses amounts to the condition that $n=2$ in 
$S_i^n=1$.

Let us mention in passing that the von Dyck groups $D(n,m,p)$ with
\be
\frac{1}{n} + \frac{1}{m} + \frac{1}{p} \leq 1 
\label{infinite}
\ee
have infinite order. Besides those listed  in Eq.~\eqref{finite}, this 
inequality leaves only the dihedral groups $\mbf{D}_p = D(2,\,2,\,p)$ as finite von Dyck groups. We will not study the dihedral groups in this paper. They have been less used for model building because they do not have irreducible representations of dimension 3. 
On the other hand, finite subgroups of the infinite von Dyck groups can be generated if one imposes additional relations to those in Eq.~\eqref{rel4}. The formalism developed in this paper will apply to those as well. The possibility of using infinite discrete groups to generate patterns of mixing will be explored elsewhere \cite{prep}.

\section{Symmetries and constraints on mixing} \label{arbitrary}

For a given set  $\{S_i$, $T_\alpha\}$ 
and a given value of $p$, the group relation, Eq.~\eqref{rel2},   provides the constraints on the 
mixing matrix $U_{PMNS}$.   
These constraints can be  found in the following way. From Eq.~\eqref{rel2}, the eigenvalues 
$\lambda_i$ of the unitary matrix
\be
W_{i\alpha U}^{-1} = U_{PMNS}S_iU_{PMNS}^\dagger T_\alpha
\ee
must be such that  $\lambda_i^p = 1$. That is, Eq.~\eqref{rel2} holds if and only if all solutions 
of the equation 
\be
\trm{Det}[W_{i\alpha U} - \lambda \mbb{I}] = 0 
\label{eig-eq}
\ee
satisfy
\be
\lambda_i^p=1 \,. \label{pthroots}
\ee 
For any unitary matrix $U_{PMNS}$ with  determinant 1,  Eq.~\eqref{eig-eq} is a cubic equation of 
the form\footnote{$W_{i\alpha U}$ is a  matrix of $SU(3)$. If its characteristic equation is 
\be
\lambda^3 + a\lambda^2 + b\lambda + c =0.    
\nn
\ee
Then 
\be
a= -(\lambda_1+\lambda_2+\lambda_3)\,,\quad b = 
\lambda_1\lambda_2+\lambda_2\lambda_3+\lambda_3\lambda_1\,,
\quad c=-\lambda_1\lambda_2\lambda_3=-\trm{Det}[W_i]=-1 \nn
\ee
and therefore 
\be
b^* = b^*\lambda_1\lambda_2\lambda_3  = \lambda_1+\lambda_2+\lambda_3 = -a,  \nn
\ee
where we have used that the eigenvalues of a unitary matrix 
are phase factors.}
\be
\lambda^3 + a\lambda^2 - a^*\lambda - 1 = 0,  
\label{cubic-gen} \,
\ee
where $a$ is a complex function of the angles and phases of $U_{PMNS}$.

From  Eq.~\eqref{eig-eq}, $a$ is given by
\be
a  = -\tr{W_{i\alpha U}}.   
\label{for-a}
\ee
If $a$ is known, this equation can be considered as a condition on  
$W_{i\alpha U}$, and consequently,  $U_{PMNS}$. This  gives two constraints 
on the  parameters of $U_{PMNS}$ which correspond to the real and 
imaginary part of $a$. 

We find the conditions on $U_{PMNS}$ assuming  for simplicity that one of the eigenvalues of 
$W_{i\alpha U} $ is equal to the identity, e.g. $\lambda_1 = 1$, which 
implies  that $a$ is real according to Eq.~\eqref{cubic-gen}.  
Then, substituting $S_i$ and $T_\alpha$ in Eq.~\eqref{for-a}
we find the general conditions on the values of 
the mixing matrix elements
\begin{align}
|U_{\beta i}|^2 &  = |U_{\gamma i}|^2\, ~~~  
\label{main1}
\\
|U_{\alpha i}|^2 & = \eta, ~~~~   \beta,\,\gamma \neq \alpha, 
\label{main2}
\end{align}
where 
\be
\eta  \equiv \frac{1-a}{4 \sin^2\left( \frac{\pi k}{m} \right)}. 
\label{eta}
\ee

The parameter $a$ can be found as the sum of the three eigenvalues $\lambda_i$  (see footnote) which, along with the previous assumption, reduces to
\be
a = - 1 - \lambda_2 - \lambda_3.
\ee 
Since $W_{i\alpha U}$ is a matrix of $SU(N)$,  we must have $\lambda_2\lambda_3 = 1$. This 
determines that ${\rm Im}[a] = 0$ and, supplemented  with $\lambda_2^p = \lambda_3^p = 1$, it 
implies that $a$ is uniquely fixed for  $p\leq 4$.  For larger $p$ two or more possibilities appear 
which should be taken into account (see below).
  
The parameter $\eta$ contains information about the flavor symmetry group. 
Eqs.~\eqref{main1} and \eqref{main2} have solutions for $\eta \leq 1$, or explicitly 
\be
\frac{1-a}{4} \leq \sin^2\left( \frac{\pi k}{m} \right) \,,
\ee
and describe  the constraints imposed on the mixing 
matrix by a flavor group $G$ defined  by the residual  symmetries $S_i$ and $T_\alpha$ of 
the neutrinos 
and charged leptons. Notice that Eq.~\eqref{main1} does not depend on $m$ or $k$. 

Notice that 
$S_i$ has a diagonal submatrix that is proportional to the identity. 
This implies that the mixing matrix that satisfies  Eq.~\eqref{for-a} is fixed at most up to a 
$2\times 2$ rotation.
On the other hand, the symmetry imposes conditions for the elements of  a single column of the mixing matrix 
(determined by $i$), Eq.~\eqref{main1}. It leads to the  equality of the elements that correspond to 
the charged leptons that transform non-trivially  under $T$. 
From  experimental data it follows that Eq.~\eqref{main1} can be satisfied either 
for $\alpha = e$, $i=1,\,2,\,3$ or for $i=2$, $\alpha = e,\,\mu,\,\tau$. That is, experimental data  implies that $T_e$ or/and $S_2$ are generators of $G$.
We consider first the case in which $T_e$ is a generator. 

We will write Eqs.~\eqref{main1} and \eqref{main2} explicitly 
in terms of the mixing angles and the CP phase using the standard  parametrization 
\be
U_{PMNS} = \left( \begin{array}{ccc}
c_{12}c_{13} & -s_{12}c_{13} & e^{-i\delta}s_{13} \\
s_{12}c_{23}+e^{i\delta}c_{12}s_{13}s_{23} & c_{12}c_{23}-e^{i\delta}s_{12}s_{13}s_{23} & -c_{13}s_{23} \\
s_{12}s_{23}-e^{i\delta}c_{12}s_{13}c_{23} & c_{12}s_{23}+e^{i\delta}s_{12}s_{13}c_{23} & c_{13}c_{23}
\end{array} \right) \,. \label{st-par}
\ee
Taking $S_1$ as a generator of $G$ and setting $\alpha =e$ we obtain
\begin{align}
\tan2\theta_{23} & = - \frac{\sin^2\theta_{12}-\cos^2\theta_{12}\sin^2\theta_{13}}{\sin 
2\theta_{12}\sin\theta_{13} \cos\delta} 
\label{th23fromS1}\\
\cos^2\theta_{12} & = \frac{\eta}{\cos^2\theta_{13}} \,. 
\label{th12fromS1}
\end{align}
For $S_2$ we have 
\begin{align}
\tan2\theta_{23} & = \frac{\cos^2\theta_{12}- \sin^2\theta_{12}\sin^2\theta_{13}}{\sin 
2\theta_{12}\sin\theta_{13} \cos\delta} \,, 
\label{th23fromS2}\\
\sin^2\theta_{12} & = \frac{\eta}{\cos^2\theta_{13}} \, . 
\label{th12fromS2}
\end{align}
The relations between $\theta_{12}$ and $\theta_{13}$, 
Eqs.~\eqref{th12fromS1} and \eqref{th12fromS2}, do  not depend on $\theta_{23}$ or $\delta$ 
while Eqs.~\eqref{th23fromS1} and \eqref{th23fromS2} yield the relation
$\tan2\theta_{23} \propto (\cos \delta \sin\theta_{13})^{-1}$.  The latter implies, in particular, 
that 2-3 mixing becomes maximal,  $\sin^2 \theta_{23} = 0.5$, in  the limits $\delta \rightarrow 
\pi/2$ and $\theta_{13} \rightarrow  0$. 
For $\delta = 0,\, \pi$ one obtains maximal deviation from maximal mixing. 
Notice that all these limits are independent of the value of $k$, $m$ and $p$.

We can substitute $\theta_{12}$ in Eqs.~\eqref{th23fromS1} and \eqref{th23fromS2} using Eqs.~\eqref{th12fromS1} and \eqref{th12fromS2}. This gives a relation between  $\theta_{23}$ and $\theta_{13}$ 
\begin{equation}
\tan2\theta_{23}  = \mp  \frac{1}{2 \sin\theta_{13} \cos\delta}  
\frac{\cos^2\theta_{13}- \eta ( 1 + \sin^2\theta_{13})}
{\sqrt{ \eta \cos^2\theta_{13} - \eta^2}}, 
\label{th23forboth}
\end{equation}
where the minus sign (plus) refers to the case $S_1$ ($S_2$). Thus, the expressions for $\tan2\theta_{23}$ differ by the sign which amounts to a substitution $\delta \rightarrow \pi - \delta$.

Eqs.~\eqref{th23fromS1} and \eqref{th23fromS2} coincide  with relations obtained in 
\cite{Ge:2011qn,Ge:2011ih,Ge:2010js}. 
In those papers, the invariance of the neutrino mass matrix  in the flavor  basis with respect to 
transformations $G_i$ was assumed. These $G_i$ in \cite{Ge:2011qn,Ge:2011ih,Ge:2010js} 
correspond to our $S_{iU} = U_{PMNS}S_iU_{PMNS}$ with the equality of  
Eq.~\eqref{main1} imposed on $U_{PMNS}$ for $\alpha = e$.   
Then, the relations between mixing angles were found using   
Eq.~\eqref{ST-transforms1}. 
Our relations have been obtained from a general group theoretical consideration
without assuming a certain form for the transformations.   
Furthermore, symmetries of the charged lepton mass matrices are included and, 
in fact, we obtain two equations which should be satisfied simultaneously to 
have a consistent group embedding. Our analysis is more general and 
can be applied to wide class of symmetries.

Choosing $S_3$ as a generating element yields
\be
\sin^2\theta_{23} =  \cos^2\theta_{23} \quad, ~~~ \quad \sin^2\theta_{13} = \eta 
\label{S3}, 
\ee
which reproduces  the well known fact that $S_3$ leads to maximal 2-3 mixing. 
For $p=2$ one has $a=1$ which gives $\sin^2\theta_{13}= 0$. 
For $p = 3,\,4$ one obtains the value $\sin^2\theta_{13} \geq 0.25$.

\vtw
On the other hand,  the cases of $p=3,\, 4$ are relevant, 
for both $S_1$ and $S_2$, in connection to models that have been  
explored extensively and we analyze them in turn. 
The models below are characterized   by the choice  of the neutrino generator $S_i$ and by numbers 
$p$ and 
$m$  which fix the charged lepton  generator and  the relation linking the two generators. We 
restrict to the cases with $k = 1$. For $m > 4$, larger values of $k$ will lead to different models.

\vsx
\noindent
{\it Model 2T} \footnote{We will denote the model by $i M$, where the number 
$i = 1,2$ corresponds to selected generator $S_i$ and the letter 
$M = B, T$ indicates the type of mixing one obtains in the limit 
$\theta_{13} = 0$: $BM  (B)$ or $TBM (T)$.}: This model uses the generator $S_2$ and $(p,m) = (3,3)$ which corresponds to  
the group $G = D(2,\,3,\,3)= {\bf A}_4$  \cite{A4tbmS2}. 
For value $p=3$ we have  
$\lambda^3 = 1$. This leads, according to Eq.~\eqref{cubic-gen}, to 
$a = 0$ and therefore  $\eta = 1/3$.  
Then from Eq.~\eqref{th12fromS2} we have
\be
\sin^2\theta_{12} = \frac{1}{3 \cos^2\theta_{13}} \,.  
\label{A4}
\ee
TBM is a point on the curve determined by this equation 
with $\theta_{13}=0$ and  $\sin^2\theta_{12}=1/3$. 
A larger absolute value of $\theta_{13}$ pulls $\theta_{12}$ 
up which is disfavored by data. 
The second relation is  as in Eq.~(\ref{th23fromS2}). 
From Eq.~(\ref{th23forboth}) we obtain 
a relation between the 2-3 and 1-3 mixings: 
\be
\tan2\theta_{23} = \frac{1 - 2 s_{13}^2}{\cos \delta  s_{13}  
\sqrt{2 - 3 s_{13}^2}}. 
\label{eqqq}
\ee

Choosing $S_1$ leads, according to Eq.~\eqref{th12fromS1}, to 
the experimentally excluded value $\sin^2\theta_{12} = 2/3$.

\vsx
\noindent
{\it Model 2B}: Uses $S_2$;  $(p,m) = (3,4)$, and consequently, the group is ${\bf S}_4$.  
We  obtain from  Eq.~(\ref{th12fromS2})
\be
\sin^2\theta_{12} =  \frac{1}{2 \cos^2\theta_{13}} \,.
\ee
It corresponds to BM mixing for $\theta_{13} = 0$. 
Hence, $\sin^2\theta_{12} \geq 0.5$ which is excluded by experimental data 
unless there are large corrections from residual symmetry breaking.
The second relation is given by Eq.~(\ref{th23fromS2}).

\vsx
\noindent
{\it Model 1B}: $S_1$,  $(p,m) = (3,4)$, {\it i.e.}  the group ${\bf S}_4$.  
Here we have the relation 
\be
\cos^2\theta_{12}  =  \frac{1}{2\cos^2\theta_{13}}\,
\ee
which contains the BM prediction \cite{S4bmS1}. 
 For allowed values of 1-3 mixing, we have from this equation $\sin^2\theta_{12} \sim 0.5$ so that  large corrections are required to make this case experimentally viable. Now the second relation is given by  Eq.~(\ref{th23fromS1}).

\vsx
\noindent
{\it Model 1T}: $S_1$,  $(p,m) = (4,3)$,  the group 
$G = D(2,\,3,\,4)= {\bf S}_4$~\cite{S4tbmS1}. 
The value $p=4$ leads to $a= -1$ so that $\eta = 2/3$ , and consequently, 
\be
\cos^2\theta_{12}  = \frac{2}{3 \cos^2\theta_{13}} \,.  
\label{S4a}
\ee
TBM corresponds to $\theta_{13}=0$, $\cos\theta_{12} = 
\sqrt{2/3}$. 
From Eq.~(\ref{th23forboth}) we obtain
\be
\tan2\theta_{23} = -  
\frac{ 1 - 5 s^2_{13}}{2\cos\delta s_{13} \sqrt{2(1 -  3 s^2_{13})}} \,.
\label{th23zzz}
\ee

\vsx
\noindent
{\it Model GR:} $S_1$,   $(p,m) = (3,5)$, the group is ${\bf A}_5$.
For $\theta_{13}=0$, Eq.~(\ref{th12fromS1}) becomes
\be
\sin^2\theta_{12} = \frac{2}{5+\sqrt{5}},  
\ee
which  reproduces the value of the Golden Ratio model~\cite{Feruglio:2011qq}.

\begin{figure}[t]
\begin{center}
\includegraphics[width=7.5cm]{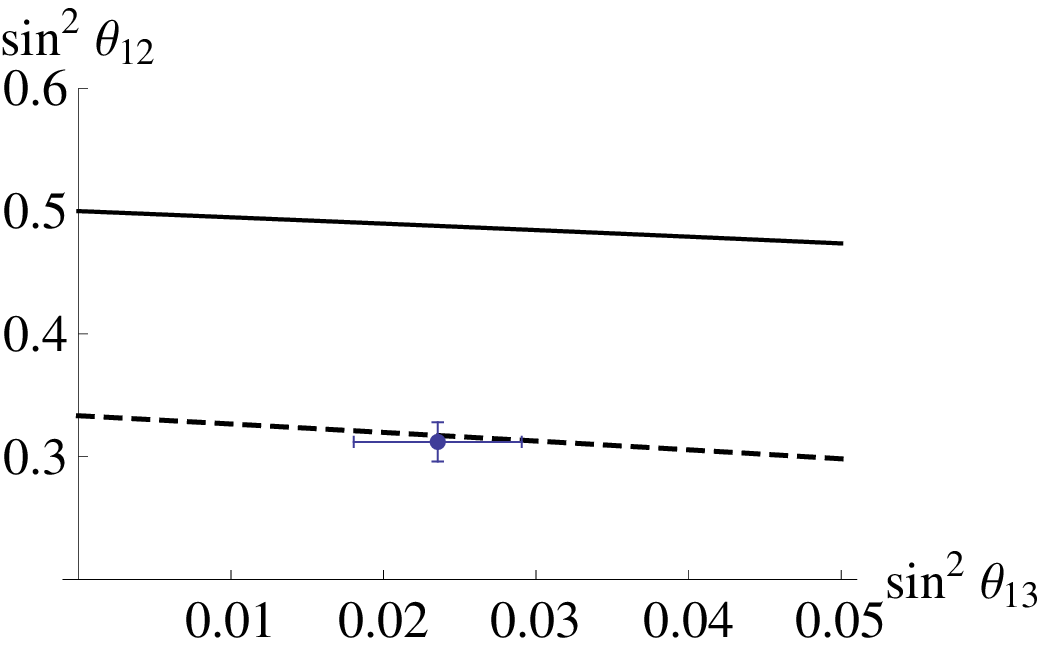}
\includegraphics[width=7.5cm]{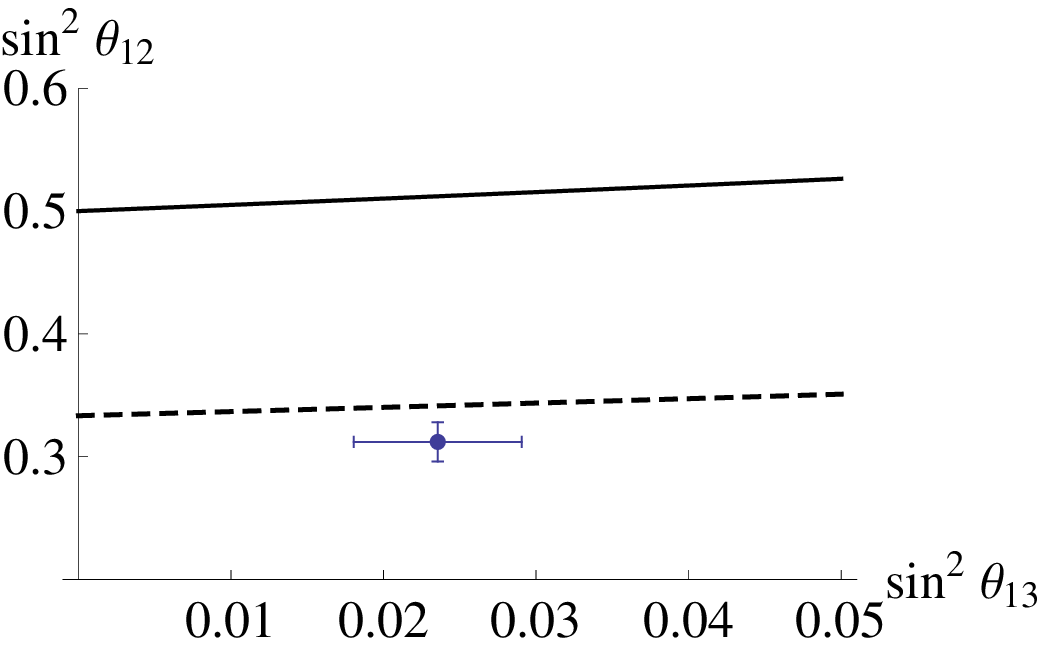}
\includegraphics[width=7.5cm]{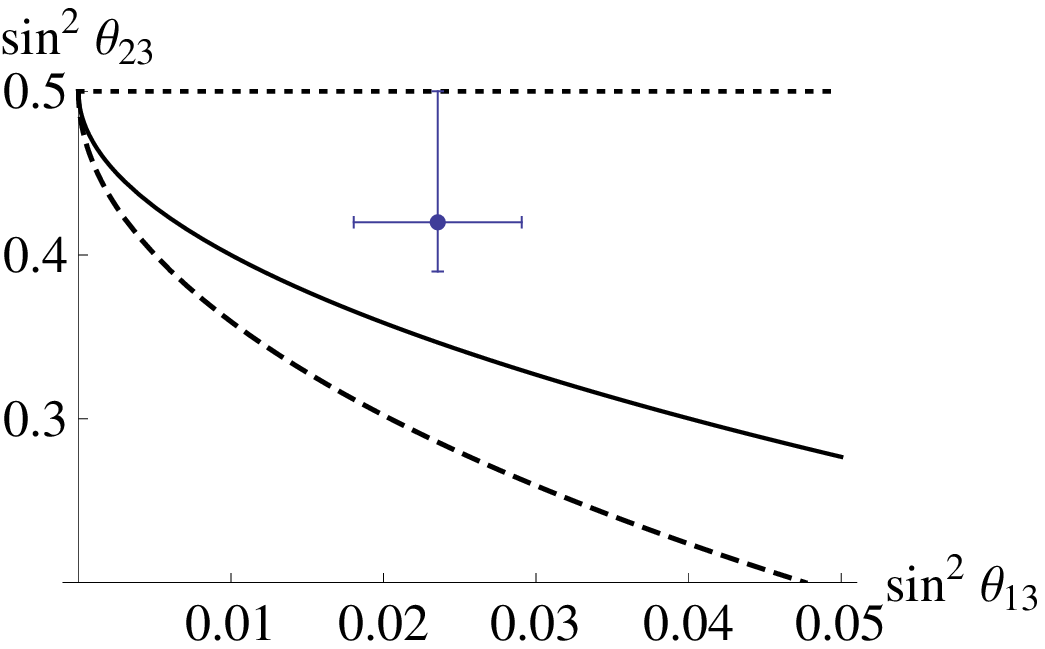}
\includegraphics[width=7.5cm]{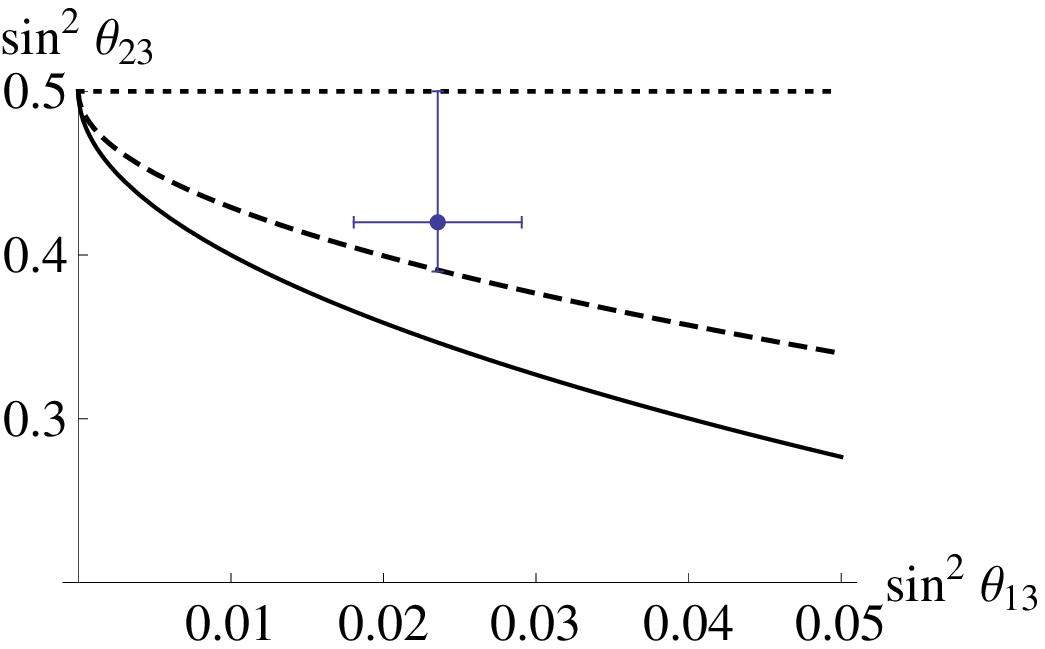}
\caption{Symmetry relations between mixing  parameters. Shown are  $\sin^2\theta_{12}$ and 
$\sin^2\theta_{23}$ as functions  of  $\sin^2\theta_{13}$ in models 1B (left, solid),  1T (left, 
dashed)  for $\delta = \pi$, 2B (right, solid), 2T (right,  dashed) for $\delta = 0$. The 
dotted line in the  lower panels  represents $\delta = \pi/2$. For  $\pi/2<\delta<\pi$ 
($0<\delta<\pi/2$) the curve is  inside the region  delimited by these two lines for the left 
(right) panels. Values of  $\delta \leq (\geq) \pi/2$ correspond to $\theta_{23}>\pi/4$. The 
crosses show the 1$\sigma$  experimentally allowed regions for the mixing angles.}
\label{fig1}
\end{center}
\end{figure}

In Fig.~\ref{fig1} we plot $\theta_{12}$  and $\theta_{23}$ as functions of $\theta_{13}$ for the models $1T$, $1B$, $2T$ and $2B$. 
The solid lines in the bottom panels delimit the predicted regions from below.
They correspond to $\delta = \pi$ for models $1T$ and $1B$ and to $\delta = 0$ for models $2T$ and $2B$. 
The panels depict regions that correspond to $\sin^2 \theta_{23} \leq 0.5$.  
The predicted regions are extended to $\sin^2 \theta_{23} > 0.5$  
symmetrically with respect to $\sin^2 \theta_{23} = 0.5$ (dotted horizontal lines). These upper parts (not shown) would correspond to the ranges $0 < \delta < \pi/2$ ($1T$ and $1B$) and $\pi/2 < \delta <  \pi$ ($2T$ and $2B$). The leftmost points  in these curves correspond to the TBM and BM values. We 
emphasize that in the lower panels of Fig.~\ref{fig1},  all the points above  the curves 
can be reached by allowing for values of $\delta \neq 0$.  
We show also best fit points and 1$\sigma$ ranges for the mixing parameters obtained from the 
global fit \cite{fit} and result of the Daya Bay experiment \cite{An:2012eh} which are in agreement with other measurements of 1-3 mixing \cite{T2K, DC, MINOS, RENO}. 
We notice that model 1T  fits 
well all the measured values for the mixing angles. For this model,  the values of $\sin^2\theta_{23}$ and $\sin^2\theta_{13}$ correspond  to  $90^\circ < \delta < 119^\circ$ with a central value of $\delta = 103^\circ$.

Since the flavor symmetry $G$ is assumed not to be an exact symmetry of the whole Lagrangian the residual symmetries are broken at the sub-leading order. 
The effect of this breaking is, however,  model-dependent and therefore can not be traced in 
our approach. The breaking can modify  the predictions presented above. Still there is certain sense 
to confront them with experimental  results: (i) corrections can be small; (ii) they can improve  
agreement with data; (iii) requirement that corrections are small can further restrict models.

In \cite{lam} it was shown that the minimal group  leading to TBM that includes \emph{all} 
symmetries of the neutrino  mass matrix is ${\bf S}_4$. 
This fact is consistent with our formalism. Indeed, if all $S_{iU}$ belong to the $G$, then
\be
(S_{iU}T_U)^{p_i} = \mbb{I}, 
\ee
so we are looking for a simultaneous solution  to all the  Eqs.~(\ref{th23fromS1})- (\ref{S3}) with 
not necessarily equal $a_i$ such that all  mixing angles have  the TBM values. 
This is easily realized for $a_1 = -1$, $a_2=0$,  $a_3=1$ corresponding to $(p_1,p_2,p_3) = 
(4,3,2)$, $k=1$, $m=3$, that is, ${\bf S}_4$. 
That is, for TBM and the values of $a_i$, $p_i$, $k$ and $m$ above, $S_2$ and $S_3$  belong to the group generated by $S_1$ and $T$. 

\vsx
Let us now consider the  cases when the group is  generated by $S_2$ and 
$T_\beta$, with $\beta = \mu$ or $\tau$. For  $T_{\tau(\mu)}$ the relations read
\begin{equation}
|U_{e 2}|^2 = |U_{\mu (\tau) 2}|^2\,,~~~~ |U_{\tau (\mu) 2}|^2 = \eta \,.
\label{mainmurau}
\end{equation}
Eliminating terms which contain the phase $\delta$  we obtain from these equations in both cases  
to the same relation  
\be
\sin^2\theta_{12} = \frac{1 - \eta}{2 \cos^2\theta_{13}} \,.
\label{th12S2}
\ee
As before, the second relation does not depend on $k$ or $m$ but  involves the CP phase $\delta$.  
For $T_\mu$ we have
\be
\cos\delta = - 2 \frac{\sin^2 \theta_{12}(\cos^2 \theta_{23} \sin^2\theta_{13} - 
\cos^2 \theta_{13}) + 
\cos^2 \theta_{12} \sin^2 \theta_{23} }{\sin 2\theta_{12} \sin 2 \theta_{23} \sin \theta_{13}}, 
\ee
while for $T_\tau$ the corresponding expression becomes
\be
\cos\delta = 
2 \frac{\sin^2\theta_{12}(\sin^2\theta_{23} \sin^2\theta_{13} - 
\cos^2 \theta_{13}) + 
\cos^2 \theta_{12} \cos^2 \theta_{23}}{\sin 2\theta_{12} \sin 2 \theta_{23} \sin \theta_{13}} 
\,.
\ee
In the limit  $\sin^2\theta_{13} \rightarrow 0$ the 2-3 mixing  converges as 
$\sin^2\theta_{23} \rightarrow  \tan^2 \theta_{12}$ 
in the $T_{\mu}$-case and $\sin^2\theta_{23} \rightarrow  
1 - \tan^2 \theta_{12}$ in  the $T_{\tau}$-case.

Notice that for $a = 0$ we  can have $|U_{e2}|^2 =|U_{\mu2}|^2 =|U_{\tau2}|^2 $ ($k=1$ and 
$m=3$), both for $T_\mu$ and $T_\tau$, which leads to the model $2T$ analyzed previously. 
An excellent  agreement with data can be obtained, for instance, for $p = 5$, $m = 6$ and $k = 1$, 
if $a = -1 + 2 \cos \pi/5 \approx 0.618$ is selected among different possibilities.  
This leads to $\eta = 0.382$. However such a choice of $p$ and $m$ corresponds to an infinite group, 
see Eq.~(\ref{infinite}).

\begin{figure}[t]
\begin{center}
\includegraphics[width=7.5cm]{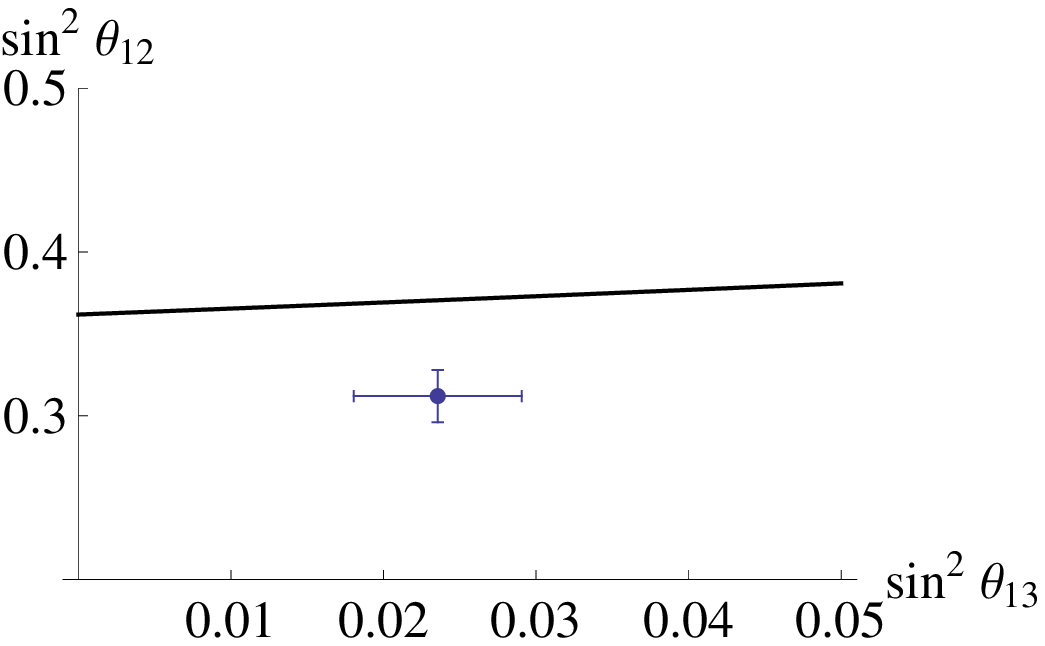}
\includegraphics[width=7.5cm]{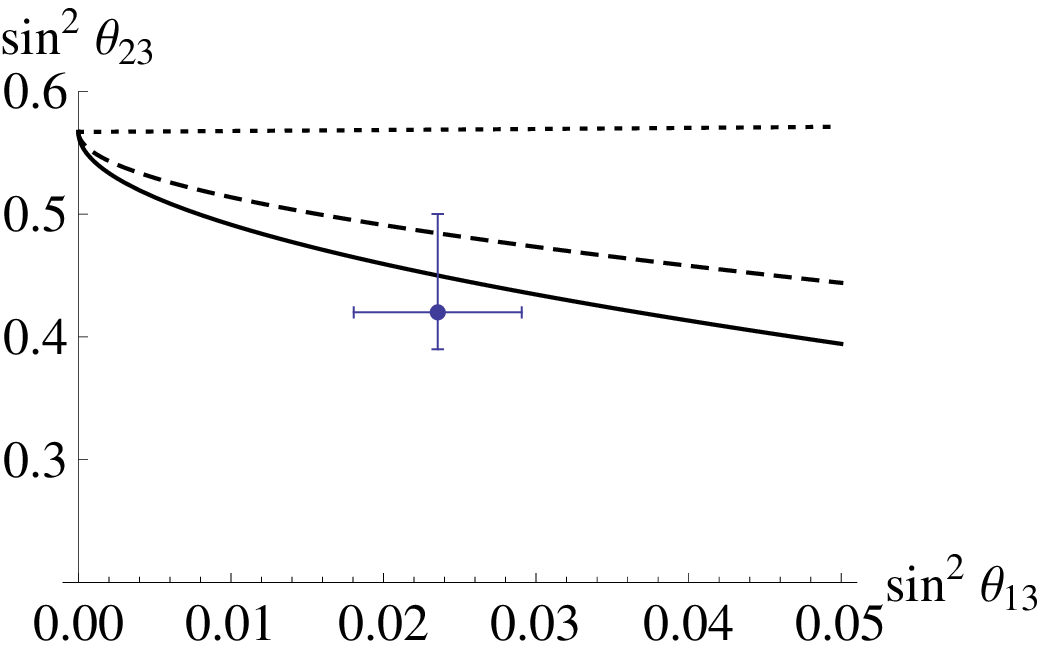}
\caption{Symmetry relations between mixing  parameters. Shown are $\sin^2\theta_{12}$ (left panel) 
and  $\sin^2\theta_{23}$ (right panel) as  functions of $\sin^2\theta_{13}$ for the model with  
$\{T_\mu;\,a=0,\,k=2,\,m=5 \}$. The curves  in the right panel correspond to $\delta = 0$ (thick), 
$\pi/4$ (dashed),  $\pi/2$ (dotted).}
\label{fig2}
\end{center}
\end{figure}

In Fig.~\ref{fig2}  we show the symmetry relations  for the model with  $T_\mu$ and $\{a,\,k,\,m \} = \{0,\,2,\,5\}$  which leads to an $\mbf{A}_5$ group ($p=3$).  The lines in the right  panel correspond to different values of the CP phase $\delta$. They move up with increasing $\delta$ and the upper line is for  $\delta = \pi/2$. In this case, the experimental measurements select small values of the CP phase: $\delta < 30^{\circ}$.

\section{Discussion and conclusions} \label{discussion}


In this paper we considered the framework in which lepton mixing appears as a result of different ways of breaking the underlying flavor symmetry in the neutrino and charged lepton Yukawa sectors. Focusing on the case of Majorana neutrinos, we developed a formalism that can be applied to various group presentations and opens up a way  of ``symmetry building'' and explanation of the data  with  discrete symmetries.

We assumed that the flavor group is determined by the residual symmetries 
of the mass matrices and considered a minimal, and at the same  time general, case in which 
only a cyclic group from each sector (the neutrino and charged lepton)  is a subgroup of the full 
flavor group $G$. A relation between the generators of these two groups  that completes the 
presentation is imposed.  The remaining symmetries of the mass matrices that do not belong to $G$ 
are treated as accidental ones.  We showed that in general, the flavor group that emerges from this 
construction is of the von Dyck type.

In this framework, we explored to which  extent  the symmetry determines the mixing. 
We derived a non-trivial matrix equation which constrains the mixing matrix 
in terms of the symmetry transformations of the neutrino and  charged leptons in the mass basis. 
These relations depend on the charges  of neutrinos and charged leptons under  the residual 
symmetries and are independent of the  basis in which the fermion fields are written. The solution 
of the matrix  equation is reduced to  2 equations for the mixing parameters. Therefore, in the 
minimal scenario in which the residual  symmetries in each sector are cyclic groups, the symmetry 
group imposes 2 relations between the 4 parameters (mixing  angles and CP phase). 
This result is compatible with  the well known fact that if only one $\mbf{Z}_2$ symmetry is imposed 
in the neutrino sector then at least one parameter of the mixing matrix must remain undetermined. 

We showed that the symmetry group allows  for non-zero 1-3 mixing and deviation from 2-3 maximal 
mixing  even in the limit of exact  residual  symmetries. With respect to TBM and BM, this paper 
clarifies  what is actually ``special'' about  them. They  correspond to limiting cases 
($\theta_{13}\rightarrow 0$)  in the continuous set of values allowed by the symmetry. We showed that, in general, 
discrete symmetries impose maximal 2-3 mixing, if either $\theta_{13} = 0$ or $\delta = \pi/2$.

The mixing can be fixed completely 
in models in which the full Klein symmetry of the neutrino 
mass matrix belongs to the flavor group. These models are particular 
cases of the framework developed here  in the sense that the larger symmetry leads to new  relations 
in addition to the ones derived in this paper. 
In the case of the flavor group $D(2, 3, 4) = {\bf S}_4$, TBM is indeed compatible with the four relations that need to hold if the Klein symmetry of the neutrino mass matrix is a subgroup of $G$.

We applied our  formalism to the finite von Dyck groups. We made a 
number of assumptions that simplify our analysis and reduce 
the number of possibilities. Namely, we assumed that:  (i) generators of the group belong to 
$SU(3)$, (ii) one of the eigenvalues of $T$ is equal to one, 
(iii) one of the eigenvalues  of the matrix $W$  is equal to 1. Furthermore, we took the group relation in the simple form, Eq.~\eqref{rel2} which leads to the von Dyck groups. Nevertheless 
we find 
that many models discussed in the literature are included in our analysis. 
Removing these assumptions 
can lead to different flavor groups and, in general, to new relations between 
the mixing parameters.

We found that $\mbf{S}_4$ can lead to a mixing matrix, 
as in model  $2T$ discussed in Sec. 3,  which is in a  good agreement with the data including the 
latest value for the 1-3 mixing. In the  simplest cases of finite groups  according to the symmetry 
relations and the experimental data on   mixing angles, the CP-violating phase should be in the 
interval $\delta = (90^{\circ} - 120^\circ)$.

\end{document}